\documentclass[a4,12pt]{elsart}

\usepackage{graphicx}
\usepackage{amsmath}

\begin{document}
\begin{frontmatter}
\title{\bf Application of spectral methods for high-frequency
  financial data to quantifying states of market participants}
\author[amp]{Aki-Hiro Sato\corauthref{cor1}}
\corauth[cor1]{Corresponding author.}
\ead{aki@i.kyoto-u.ac.jp}
\address[amp]{Department of Applied Mathematics and Physics, 
Graduate School of Informatics, Kyoto University,
Kyoto 606-8501, Japan.} 

\date{\empty}

\begin{abstract}
Empirical analysis of the foreign exchange market is conducted based
 on methods to quantify similarities among multi-dimensional time
 series with spectral distances introduced in [A.-H. Sato, Physica A,
 {\bf 382} (2007) 258--270]. As a result it is found that the similarities
 among currency pairs fluctuate with the rotation of the earth, and that
 the similarities among best quotation rates are associated with those
 among quotation frequencies. Furthermore it is shown that the
 Jensen-Shannon spectral divergence is proportional to a mean of the
 Kullback-Leibler spectral distance both empirically and numerically. It
 is confirmed that these spectral distances are connected with
 distributions for behavioral parameters of the market participants from
 numerical simulation. This concludes that spectral distances of
 representative quantities of financial markets are related into
 diversification of behavioral parameters of the market participants.
\end{abstract}

\end{frontmatter}
\maketitle

\thispagestyle{empty}
\section{Introduction}
\label{sec:Introduction}
Developing and spreading of Information and Communication Technology
(ICT) computerize our society all over the world. One can access 
information more easily than before 1980s, while one sometimes 
faces to occasions when one has to handle massive information beyond 
human information processing capacity. 

In early 1990s financial markets started to introduce electronic trading
systems due to development of ICT. This brings that information
generated in the financial markets almost becomes detectable. It is
even becoming possible to conduct algorithmic trading or automated
trading by using real-time financial data.


The foreign exchange market is the largest financial market 
all over the world. Recently an increasing number of financial
institutions are embracing the opportunities and benefits in automated
and algorithmic trading systems. These systems have been 
successively studied in order to avert risks due to 
explosion of information and several financial institutions even 
experimentally or commercially utilize
them~\cite{Casqueiro,Dempster,Strozzi}. It is predicted that more becoming
common more trading can be quickly conducted. Moreover they 
have an advantage to human traders from a viewpoint of fairness,
accuracy, speed, and cost-effectiveness. If they excess critical mass 
then human traders may not understand what happens in the financial markets. 
Therefore it is expected to become an important work to construct
a system to quantify market states (states of market participants in
financial markets) automatically since monitoring of financial markets
of which speed is predicted to be beyond human information
processing capacity is needed. From physical point of view practical
methods should be considered. In order to construct such a system 
constructive utilization of high frequency financial data should be 
sophisticated. 

Generally speaking massive data contain less information since one
cannot understand and handle them without computers. In order to
understand them some kinds of information extraction techniques are
proposed~\cite{Carbone}. By application of the random matrix theory to
correlation matrices of multi-dimensional time series some studies
showed that market states reveal in both eigenvectors and
eigenvalues~\cite{Laloux:99,Tibely:06}. It is known that the bone
structure consisting of larger eigenvectors is related to clusters of
financial markets. This approach is based on methods to 
extract network structures of correlation network~\cite{Aste:06}. In
fact correlations among financial commodities are related to market
structure but multiple aspects to measure states of financial markets
should be discussed.

In this article we consider alternative method to quantify market states 
with spectral methods. Moreover the relation between spectral distance 
and behavioral parameters of agents is discussed through an agent-based
model. Recent works show that agent-based models of financial markets
are useful to understand price formation mechanism in the financial 
markets~\cite{Yamada:07,Challet}.

The rest of this article is constructed as follows. In Section
\ref{sec:spectral-distance} definition of spectral distance for
multi-dimensional time series is presented. In Section
\ref{sec:representative-quantities} quotation rates and quotation
frequencies are introduced as representative quantities of financial
markets. In Section \ref{sec:empirical-analysis} the Jensen-Shannon
spectral divergence and the Kullback-Leibler spectral distance
are calculated by using high-frequency financial data of the foreign
exchange market. In Section \ref{sec:agent-based model} in order to
check adequateness of proposal methods those spectral distances are
calculated by an agent-based model. Section \ref{sec:conclusion} is
devoted into concluding remarks.

\section{Definition of spectral distances for multi-dimensional time series}
\label{sec:spectral-distance}
Consider $M$-dimensional time series $x_j(k) \quad
(j=1,\ldots,M; k=0,1,\ldots,L-1)$ with a sampling period $\Delta t$. 
The power spectrum of $x_j(k)$ is estimated as periodgram estimator with
the Hanning windowing function with width $N \quad (<L)$,  
\begin{equation}
w(k) = \frac{1}{2}\Bigl(1-\cos(\frac{2\pi k}{N-1})\Bigr).
\end{equation}
The periodgram estimator is calculated as
\begin{equation}
P_j(f_n,t) =
 \frac{1}{N^2}\Bigl|\sum_{k=0}^{N-1}w(k)x_j(k+t)e^{-2\pi\mbox{i}k\frac{n}{N}}\Bigr|^2
 \quad \Bigl(f_n = \frac{n}{N\Delta t}\Bigr),
\end{equation}
where the Nyquist frequency is $f_c = 1/(2\Delta t)$. By using the
normalized power spectrum, which is defined as 
\begin{equation}
p_j(f_n,t) = \frac{P_j(f_n,t)}{\sum_{n=1}^{N-1} P_j(f_n,t)},
\end{equation}
the spectral entropy is calculated as its Shannon entropy;
\begin{equation}
H(p_j,t) = -\sum_{n=1}^{N-1} p_j(f_n,t) \log p_j(f_n,t).
\label{eq:spectral-entropy}
\end{equation}
Note that a direct current component is ignored in this
definition. Obviously when $x_j(k)$ is a white noise it takes the maximum
value since the power spectrum of it is a uniform  function. Contrarily
when $x_j(k)$ is a sinusoidal function it takes the minimum value since
the power spectrum of it is a Dirac's $\delta$-function.

In fact the spectral entropy indicates randomness of time
series. High/low $H$ means that $x_j(k)$ is unpredictable/predictable. 
Suppose the situation to predict time series from historical
observation. The mode for a normalized power spectrum is the frequency
where the normalized power spectrum has the maximum value:
\begin{equation}
\overline{f_j(t)} = \mbox{arg}\quad\underset{f_n}{\max}\{p_j(f_n,t)\}.
\end{equation}
If the spectral entropy is low then sinusoidal signal with $\overline{f_j(t)}$
is an adequate model to predict the future time series. Contrarily if the
spectral entropy is high then sinusoidal signal with $\overline{f_j(t)}$ is
unworkable to predict the future time series. Therefore the spectral
entropy is related into predictability of the mode on a frequency domain. 

Actually according to the maximum entropy principle the best choice for
the distribution $p_j(f_n,t)$ is the one that maximizes
Eq. (\ref{eq:spectral-entropy}) subject to the constraint the normalized
condition
\begin{equation}
\sum_{n=1}^{N-1} p_j(f_n,t) = 1.
\end{equation}
By using the Lagrange multiplier differentiating
\begin{equation}
L = -\sum_{n=1}^{N-1}p_j(f_n,t)\log p_j(f_n,t) + \lambda\Bigl(1-\sum_{n=1}^{N-1}p_j(f_n,t)\Bigr),
\end{equation}
with respect to $p_j(f_n,t)$ and then equating the
result to zero one obtains the uniform normalized power spectrum
\begin{equation} 
p_j(f_n,t)=\frac{1}{N-1}.
\end{equation}
Namely the best choice without knowledge is the uniform power spectrum,
which means that time series are white noises.

By using the spectral entropy the Jensen-Shannon spectral
distance~\cite{Lin:91,Sato:07} for $M$-dimensional time series is
defined as
\begin{equation}
JS(t) = H\Bigl(\sum_{j=1}^{M}\pi_j p_j,t\Bigr) - \sum_{j=1}^{M} \pi_j H(p_j,t),
\end{equation}
where $\sum_{j=1}^M \pi_j = 1$ and $\pi_j>0$, one can quantify the
similarity among a set of time series. The Jensen-Shannon spectral
distance measures similarity of randomness among time series. As an
alternative spectral distance the Kullback-Leibler spectral
distance defined as
\begin{equation}
KL_{lm}(t) =
 \sum_{n=1}^{N-1}p_l(f_n,t)\log\frac{p_l(f_n,t)}{p_m(f_n,t)} 
\end{equation}
can be considered in analogy with relative entropy between normalized power
spectra~\cite{Veldhuis:03,Sato:06,Sato:07}.

\section{Representative quantities of financial markets}
\label{sec:representative-quantities}
\begin{figure}[hbt]
\centering
\includegraphics[scale=0.4]{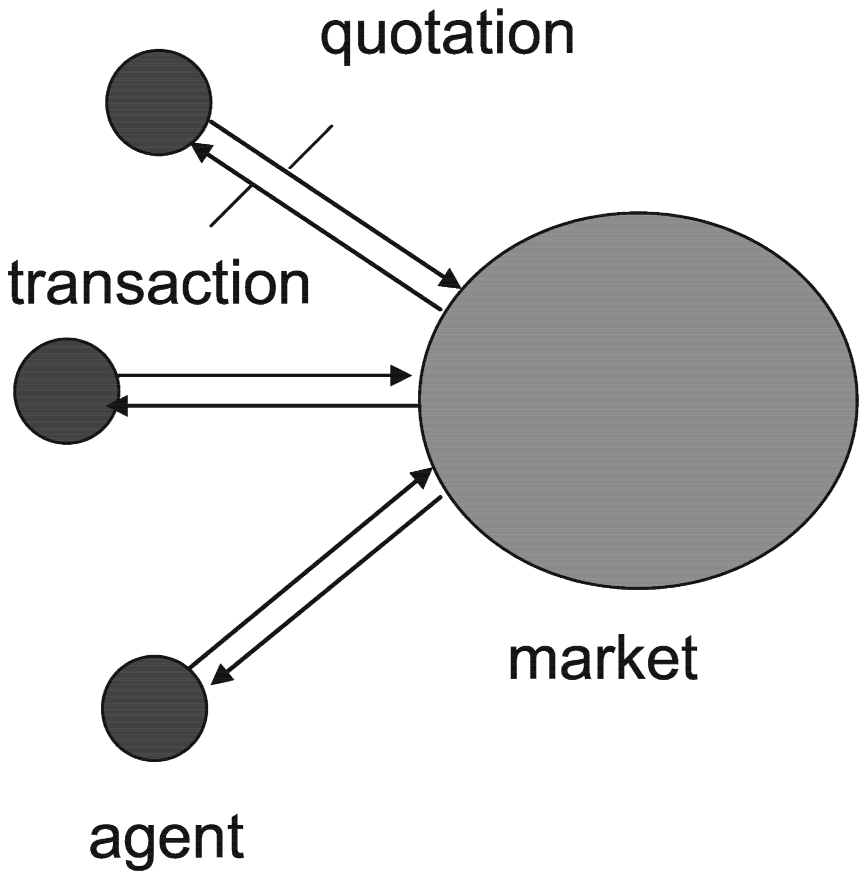}
\caption{A pictorial illustration of circular causality of a financial
 market.}
\label{fig:FX}
\end{figure}

What are representative quantities of financial markets? 
Since there is circular causality between market participants and a
broking system quotations are causes/results for broking 
systems/market participants, and transactions are results/causes
(see. Fig. \ref{fig:FX}).

For empirical investigation the database of the foreign exchange market 
in the CQG datafactory is used~\cite{CQG}. As representative quantities
of the financial market quotation rates and quotation frequencies are
detectable from this database. The data contain 20 currency pairs
(AUD/JPY, AUD/USD, CAD/JPY, CHF/JPY, EUR/AUD, EUR/CHF, EUR/GBP,
EUR/JPY, EUR/NOK, EUR/SEK, EUR/USD, NZD/USD, USD/CAD, USD/CHF,
USD/JPY, USD/NOK, USD/SEK, USD/ZAR, GBP/AUD, and GBP/JPY)
which consists of 11 currencies; AUD(Australia Dollar), CAD(Canadian
Dollar), CHF(Swiss Franc), EUR(Euro), GBP(United Kingdom Pound),
JPY(Japanese Yen), NOK(Norwegian Krone), NZD (New Zealand Dollar), 
SEK(Swedish Krona), USD(United States Dollar), and ZAR(South Africa
Rand).

Quotation frequencies are defined as the number of quotations per unit time,
\begin{equation}
A_j(k) = \frac{1}{\Delta t}C_j(k\Delta t;(k+1)\Delta t) \quad
 (k=0,1,\ldots; j=1,\ldots,20),
\end{equation}
where $C_j(t_1;t_2)$ is the number of ask/bid quotations of the $j$-th 
currency pair from $t_1$ to $t_2$, and $\Delta t$ denotes the
sampling period. $\Delta t$ is set as 1 [min] throughout this
analysis ($f_c = 0.5$ [1/min]). The best ask/bid rates are defined as
the minimum ask/maximum bid rates between $k\Delta t$ and $(k+1)\Delta t$.
\begin{equation}
R_j(k) =
\left\{
\begin{array}{ll}
\underset{k\Delta t \leq t' \leq (k+1)\Delta t}{\min}\{ask_j(t')\}/
\underset{k\Delta t \leq t' \leq (k+1)\Delta t}{\max}\{bid_j(t')\} &
 (A_j(k)\neq 0) \\
R_j(k-1) & (A_j(k) = 0)
\end{array}
\right.,
\end{equation}
where $ask_j(t)/bid_j(t)$ denotes ask/bid rates of the $j$-th currency
pair at time $t$.

In this article, on the basis of spectral methods quantifying dynamical
structure of best quotation rates and quotation frequencies is conducted.

\section{Empirical analysis of the foreign exchange market}
\label{sec:empirical-analysis}
\begin{figure}[hbt]
\includegraphics[scale=0.4]{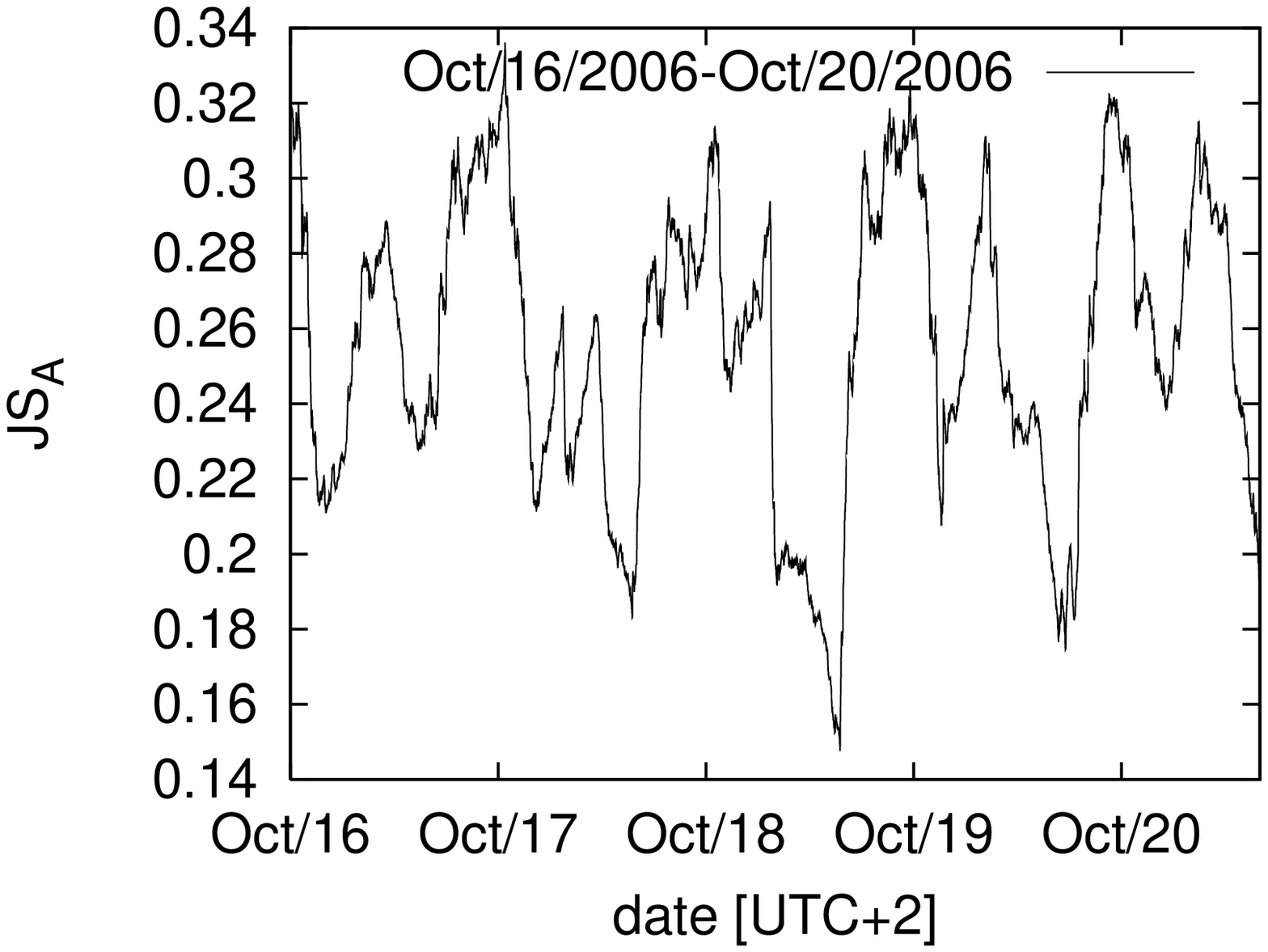}(a)
\includegraphics[scale=0.4]{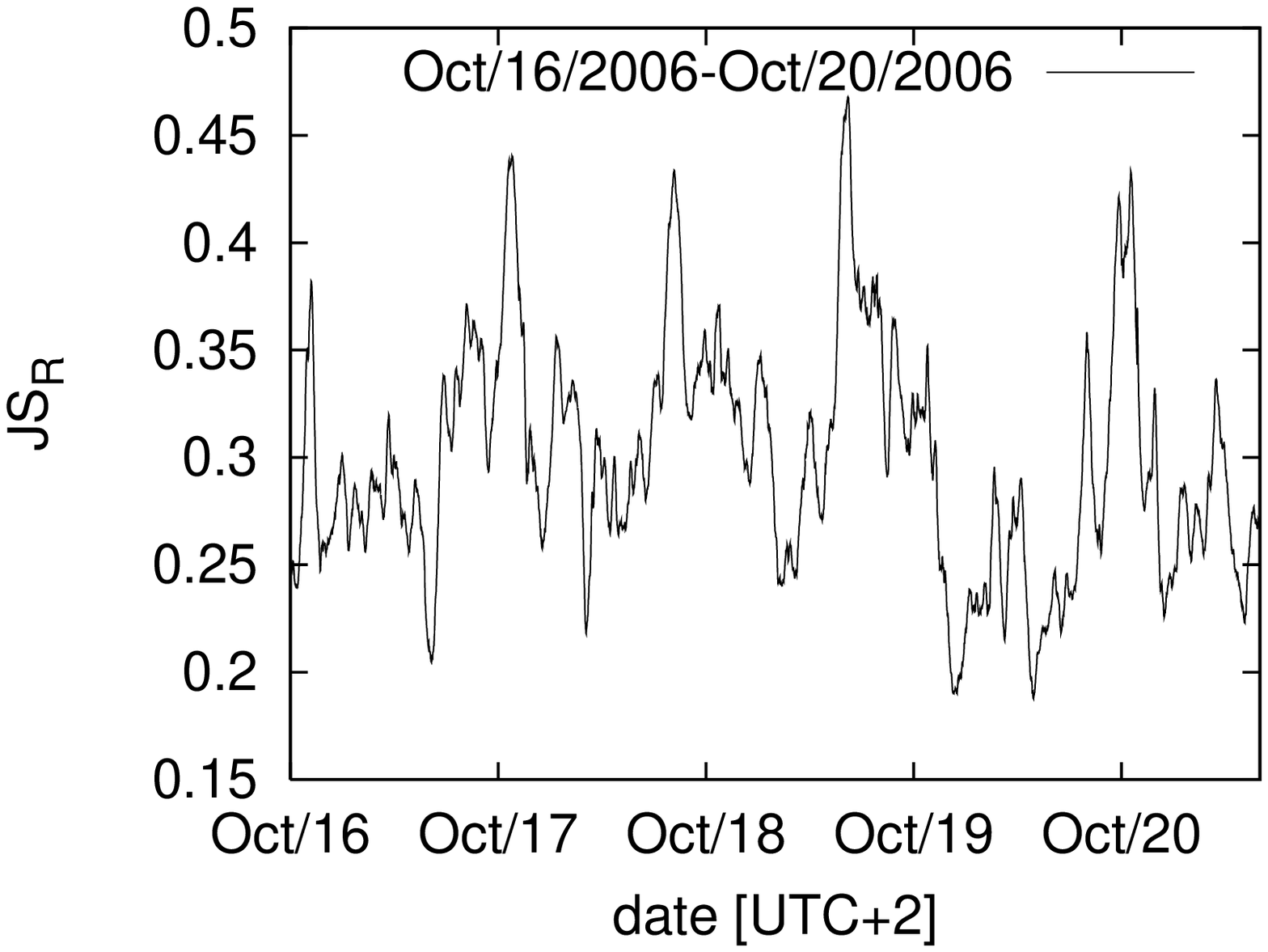}(b)
\caption{The Jensen-Shannon spectral divergence of 20 currency pairs 
for the ask quotation frequencies (a) and that for the best ask rates (b) 
for a period from 16th  October 2006 to 20th October 2006.} 
\label{fig:JSs}
\end{figure}

As shown in Fig. \ref{fig:JSs} the Jensen-Shannon spectral distances for
best ask rates and ask quotation rates fluctuate with rotation of the
earth. The Jensen-Shannon spectral divergence for the ask quotation
frequencies denoting $JS_A$ and that for the best ask rates denoting
$JS_R$ are calculated as shown in Fig. \ref{fig:JSs}. 

In order to calculate a correlation between them a cross-correlation
coefficient is introduced as  
\begin{equation}
C_{RA} = \frac{\langle JS_R JS_A \rangle - \langle JS_R \rangle \langle
 JS_A \rangle}{\sqrt{\mbox{Var}\{JS_R\}}\sqrt{\mbox{Var}\{JS_A\}}}.
\end{equation}
It is estimated as $C_{RA}=0.624966$ for a period from 16th Oct 2006 to
20th Oct 2006. As a result it is found that best ask rates and 
quotation frequencies are weekly and constantly coupled with each other.

\begin{figure}[hbt]
\centering
\includegraphics[scale=0.4]{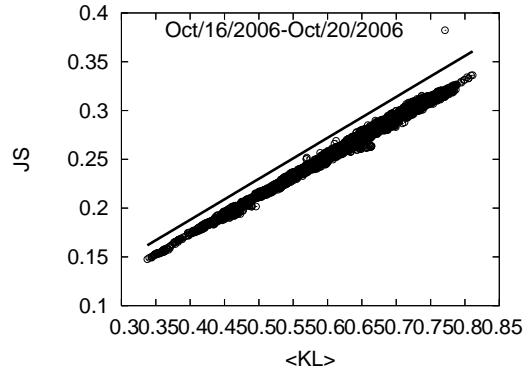}
\caption{The relation between a mean of the Kullback-Leibler
 spectral distance and the Jensen-Shannon spectral distance for 20
 currency pairs. A straight line represents a slope (0.42) obtained
 from the least square method, and circles are obtained from empirical
 data.}
\label{fig:prop}
\end{figure}

Furthermore it is confirmed that values of the Jensen-Shannon spectral
distance are related to the Kullback-Leibler spectral distance matrices.
A mean of the Kullback-Leibler spectral divergence, defined as
\begin{equation}
\langle KL \rangle(t) = \frac{1}{M^2}\sum_{l=1}^{M}\sum_{m=1}^{M}KL_{lm}(t)
\end{equation}
is always greater than $JS(t)$ ($\langle KL \rangle (t) \geq JS(t)$). 
Furthermore it is found that $JS(t)$ is proportional to 
$\langle KL \rangle(t)$ with a coefficient as 0.42 from the least square
method for empirical results as shown in Fig. \ref{fig:prop} ($JS(t)
\propto \langle KL \rangle(t)$). 

\section{Agent-based model}
\label{sec:agent-based model}
By introducing an agent-based model of a financial market
in which $N$ market participants exchange $M$ financial
commodities~\cite{Sato:07}.

The agents have threshold dynamics like the Granovetter
model~\cite{Granovetter:78}. Three investment attitudes are codes as
1 (buying), 0 (waiting), and -1 (selling). If the inner states of the 
$i$-th agent for the $j$-th financial commodity at time
$t\quad(t=0,1,2\ldots)$ excesses buying threshold $\theta_{ij}^B(t)$ or
below selling threshold $\theta_{ij}^S(t)$ then the $i$-th agent
selects the investment attitude for the $j$-th financial commodity at
time $t$
\begin{equation}
y_{ij}(t)=
\left\{
\begin{array}{ll}
1  & (\Phi_{ij}(t) \geq \theta_{ij}^B(t)) \\
0  & (\theta_{ij}^S(t) < \Phi_{ij}(t) < \theta_{ij}^B(t)) \\
-1 & (\Phi_{ij}(t) \leq \theta_{ij}^S(t))
\end{array}
\right.,
\end{equation}
where $\Phi_{ij}(t)$ is assumed to be described within diversification of
interpretation $\xi_i(t)$ and self-feeling $a_{ij}(t)$:
\begin{equation}
\Phi_{ij}(t) = a_{ij}(t)\Bigl(x_i(t)+\xi_i(t)\Bigr),
\end{equation}
where $x_i(t)$ denotes perception of the $i$-th market participants.
For simplicity the perception is assumed to be described as both
exogenous and endogenous information;
\begin{equation}
x_i(t) = \sum_{k=1}^{M}c_{ik}(|\theta_{ik}^S(t)|,|\theta_{ik}^B(t)|)\frac{1}{T}\sum_{\tau=1}^{T}r_k(t-\tau\Delta t) + s_i(t), 
\label{eq:information}
\end{equation}
where $r_j(t)=\gamma/N \sum_{i=1}^Ny_{ij}(t)$ denote log returns,
$s_i(t)$ exogenous information, $c_{ik}(\cdot,\cdot)$ functions to
show attentions, and $T (\geq 1)$ a time span to calculate a moving
average. The attentions are assumed as $C_{ik}(x,y)=1/(x^2+y^2)$.
Rates $R_j(t)$ and quotation frequencies $A_j(t)$ of the $j$-th
financial commodity at time $t$ are respectively defined as
\begin{equation}
R_j(t+\Delta t) = R_j(t)\exp\Bigl(r_j(t)\Bigr), \quad
A_j(t) = \frac{1}{\Delta t}\sum_{i=1}^N|y_{ij}(t)|
\end{equation}

\begin{figure}[hbt]
\centering
\includegraphics[scale=0.37]{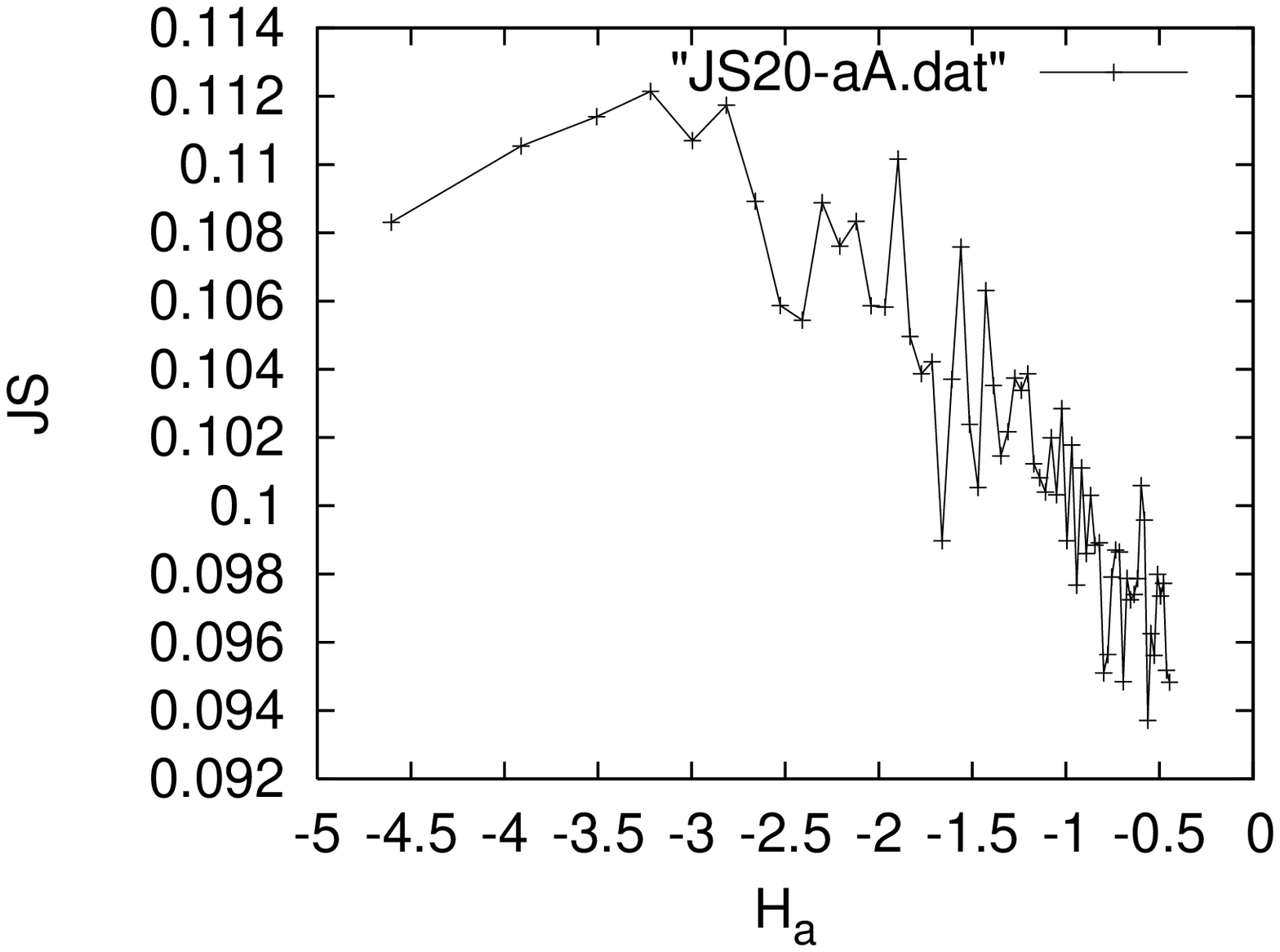}(a)
\includegraphics[scale=0.37]{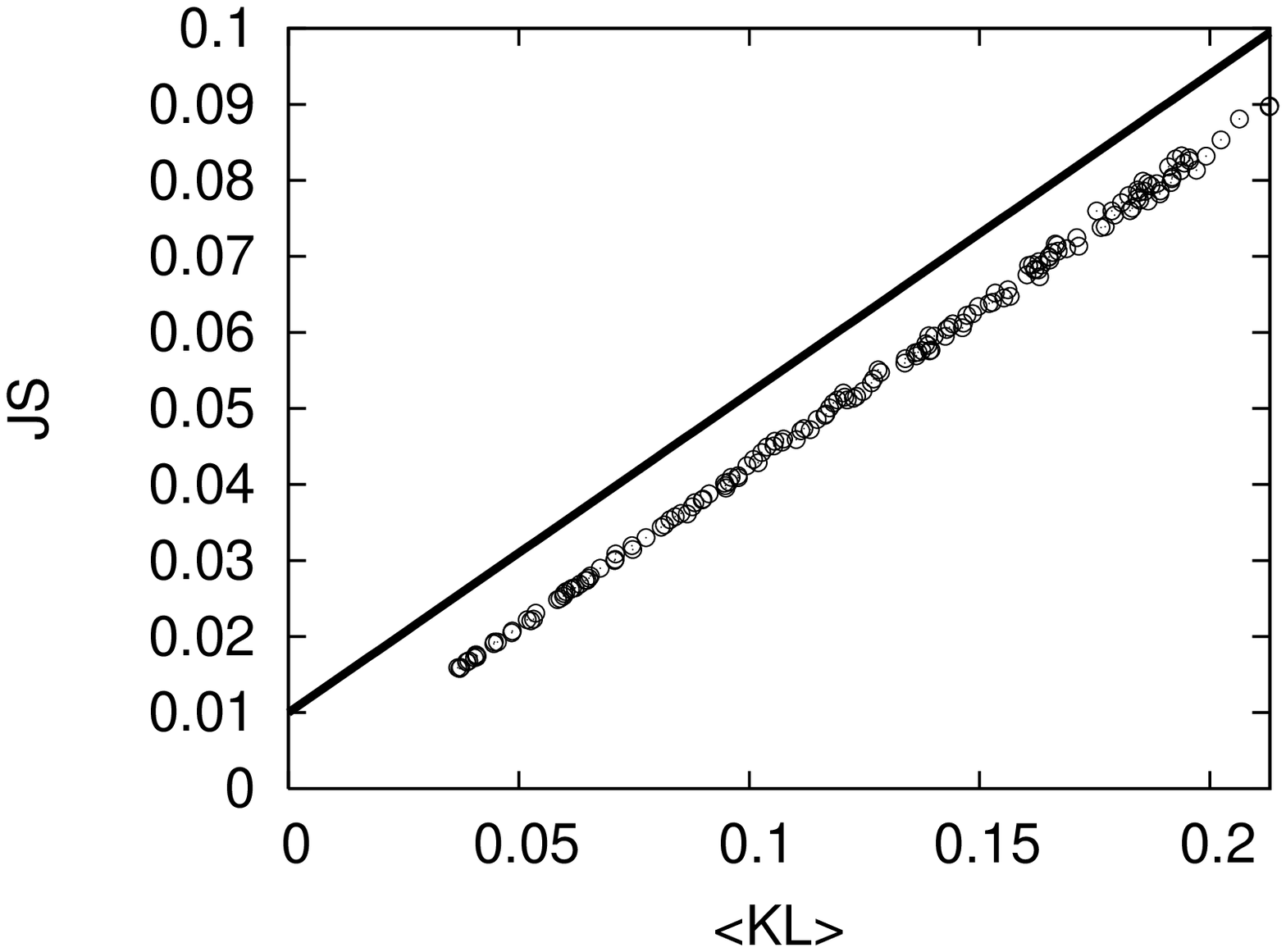}(b)
\caption{(a) The relationship between parameters and JS of behavioral
 frequencies. The diversification of $a_{ij}$ is measured as an
 entropy of parameter distribution. Parameters are fixed as $N=2000$,
 $M=20$ and $T=1$. $\theta_{ij}^S(t)$, $\theta_{ij}^B(t)$, and $a_{ij}(t)$
 are sampled from uniform distribution. The ranges of thresholds are
 fixed as $[-0.02,-0.01]$ and $[0.01,0.02]$, and the range of
 $a_{ij}(t)$ $[a_1, a_2]$ is measured by entropy $H_a = \log (a_2 - a_1)$;
 (b) The relation between a mean of the Kullback-Leibler spectral
 distance and the Jensen-Shannon spectral distance. A straight line
 represents a slope (0.42) obtained from the least square method, and
 circles are obtained from numerical simulation.}
\label{fig:simulation}
\end{figure}

Relationships between spectral distance and behavioral parameters of 
the market participants are investigated. It is found that
diversification of parameters of agents is related to the Jensen-Shannon
spectral distance for behavioral frequencies as shown in
Fig. \ref{fig:simulation} (a). Furthermore it is confirmed that 
a mean of the Kullback-Leibler spectral distance matrices is
proportional to the Jensen-Shannon spectral divergence from numerical
simulation as shown in Fig. \ref{fig:simulation} (b).

\section{Conclusion}
\label{sec:conclusion}
Spectral distances can be adopted to quantity to measure the market
states totally. As a result it is found that the spectral distances
fluctuate with rotation of the earth and that best quotation rates are
weakly coupled with quotation frequencies. The Jensen-Shannon spectral
distance is proportional to a mean of the Kullback-Leibler spectral
distance.

It is expected that employment of both spectral methods and
high-frequency financial data provides us deepen insights of financial
markets with high resolution. Information extraction from massive data
will become important to understand states of financial markets.

This work was partially supported by a Grant-in-Aid for Scientific
Research (\# 17760067) from the Japanese Ministry of Education, Culture,
Sports, Science and Technology.


\begin{thebibliography}{99}
\bibitem{Casqueiro}
	P.X. Casqueiro and A.J.L. Rodrigues,
	European Journal of Operational	Research, {\bf 175} (2006)
	1400--1412. 
\bibitem{Dempster}
	M.A.H Dempster and V. Leemans, Expert Systems with Applications, 
	{\bf 30} (2006) 543--552.
\bibitem{Strozzi}
	F. Strozzi, J.-M. Zald\'ivar, J.P. Zbilut, Physica A, {\bf 376}
	(2007) 487--499.
\bibitem{Carbone}
	A. Carbone, H.E. Stanley, Physica A, {\bf 384} (2007) 21--24.
\bibitem{Laloux:99}
	L. Laloux, P. Cizeau, J.-P. Bouchaud, and M. Potters,
	Phys. Rev. Lett., {\bf 83} (1999) 1467--1470.
\bibitem{Tibely:06}
	G. Tib\'ely, J.-P. Onnela, J. Saram\"aki, K. Kaski, and
	J. Kert\'esz, Physica A, {\bf 370} (2006) 145--150.
\bibitem{Aste:06}
	T. Aste, and T.D. Matteo, Physica A, {\bf 370} (2006) 156--161.
\bibitem{Yamada:07}
	K. Yamada, H. Takayasu and M. Takayasu, Physica A, {\bf 382}
	(2007) 340--346.
\bibitem{Challet}
	D. Challet, Physica A, {\bf 382} (2007) 29--35.
\bibitem{Lin:91} 
         J. Lin, IEEE Transaction on information theory, {\bf 37} (1991)
	145--150.
\bibitem{Sato:07} 
	A.-H. Sato, Physica A, {\bf 382} (2007) 258--270.
\bibitem{Veldhuis:03}
	R. Veldhuis, and E. Klabbers, IEEE transactions on speech and
	audio processing, {\bf 11} (2003) 100--103.
\bibitem{Sato:06} A.-H. Sato, J. Oshiro, Journal of the Physical Society
	of Japan, {\bf 75} (2006) 084005.
\bibitem{CQG}
	Data is provided by CQG Inc.
\bibitem{Granovetter:78}
	M. Granovetter, The American Journal of Sociology, {\bf 83}
	(1978) 1420--1443.
\end{thebibliography}
\end{document}